\RequirePackage[2020-02-02]{latexrelease}

\documentclass[aip,reprint,floatfix]{revtex4-1}

\usepackage{bm}% bold math
\usepackage{amsmath}% subequations
\usepackage{amssymb, mathrsfs}% frakmath
\usepackage{graphicx}% Include figure files
\providecommand{\MakeTextUppercase}[1]{\MakeUppercase{#1}}

\begin{document}

\title{Passive acoustic non-line-of-sight localization without a relay surface}
\author{Tal I. Sommer}
\affiliation{Department of Applied Physics, Hebrew University of Jerusalem, Jerusalem 9190401, Israel}
\affiliation{Alexander Grass Center for Bioengineering, Hebrew University of Jerusalem, Jerusalem 9190401, Israel}

\author{Ori Katz}
%\email[Author to whom correspondence should be addressed: ]{orik@mail.huji.ac.il}
\affiliation{Department of Applied Physics, Hebrew University of Jerusalem, Jerusalem 9190401, Israel}

\date{\today}

\begin{abstract}

The detection and localization of a source hidden outside the Line-of-Sight (LOS) traditionally rely on the acquisition of indirect signals, such as those reflected from visible relay surfaces such as floors or walls. These reflected signals are then utilized to reconstruct the obscured scene. In this study, we present an approach that utilize signals diffracted from an edge of an obstacle to achieve three-dimensional (3D) localization of an acoustic point source situated outside the LOS. We address two scenarios - a doorway and a convex corner - and propose a localization method for each of them. For the first scenario, we utilize the two edges of the door as virtual detector arrays. For the second scenario, we exploit the spectral signature of a knife-edge diffraction, inspired by the human perception of sound location by the head-related transfer function (HRTF). In both methods, knife-edge diffraction is utilized to extend the capabilities of non-line-of-sight (NLOS) acoustic sensing, enabling localization in environments where conventional relay-surface based approaches may be limited.

\end{abstract}

\maketitle
 
%##### INTRO & BACKGROUND #####
The challenge of Non-Line-of-Sight (NLOS) imaging has interesting applications in diverse fields, including autonomous navigation, remote surveillance, search-and-rescue, and more. Obtaining a direct line of sight in these scenarios may be unsafe, challenging, or even impossible. Addressing the challenge involves reconstructing the location and potentially also the shape of objects hidden from direct view, such as those around corners or behind barriers \cite{wetzstein2020NLOS}.

In the past decade, there has been a growing interest and research focus on optical and acoustical techniques, both active and passive, that perform NLOS imaging, with impressive results in a variety of scenarios. 
Both optical and acoustical-based techniques seek to address this challenge by extracting information from indirect reflected signals that reach the detection system after being either specularly- or diffusively- reflected from a visible relay surface, such as a wall, a floor, or a ceiling.
While active NLOS imaging systems, both optical\cite{velten2012NLOS3d, gariepy2016NLOS_moving, wetzstein2020NLOS, Wetzstein2018optic_conNLOS} and acoustical\cite{Wetzstein2019acousticNLOS} exploit a relay surface to reflect controlled emitted signals to the target object and back, passive systems\cite{czajkowski2024two_edges_passNLOS, freeman2017corner_cam, freeman2020_2d_1edge, Jeremy2022passive_acoustic_NLOS, Jeremy2019passive_optical_NLOS, katz2014nlos, katz2012nlos2} detect reflected signals that are emitted from the target source itself to infer location. In this manuscript, we will discuss passive localization systems.

Traditionally, NLOS localization relies on measuring the time-of-arrival (TOA) and/or intensity distribution on a visible relay surface\cite{Wetzstein2018optic_conNLOS, velten2024opticNLOS_act_pas, freeman2017corner_cam, freeman2020_2d_1edge} to determine the position and shape of targets hidden from view.
In the case of TOA measurements, every point on the relay surface effectively serves as a virtual detector, and the target can be localized in 3D from this two-dimensional 'array' of virtual detectors \cite{wetzstein2020NLOS, Velten2020phasor}.
In the case of intensity-only measurements, the shadow of the edge of the barrier is utilized to infer the source location. When the barrier has only one edge, as in a convex corner, localization information is available only in one axis \cite{freeman2017corner_cam}.

If a relay surface is not present, or if the reflections from the relay surface do not reach the detection system, e.g., as in the case of specular mirror-like reflections in acoustics, one is limited to utilizing the signals that arrive directly from the visible edges of the barrier/corner. Such 'direct' signals indeed exist, as the sharp edge of the barrier diffracts the signal from the source in whats known as "knife-edge diffraction" \cite{hum2007ked_theory}.
%In this case, signals emitted from the target source reach the edges of the barrier. The finite or semi-finite aperture (Fig.~\ref{fig:door_scene}a and Fig.~\ref{fig:ked_scene}) causes the acoustic field to diffract onto the detection system, which is outside of the direct LOS from the source.

\begin{figure*}[!ht]
    \centering
    \includegraphics{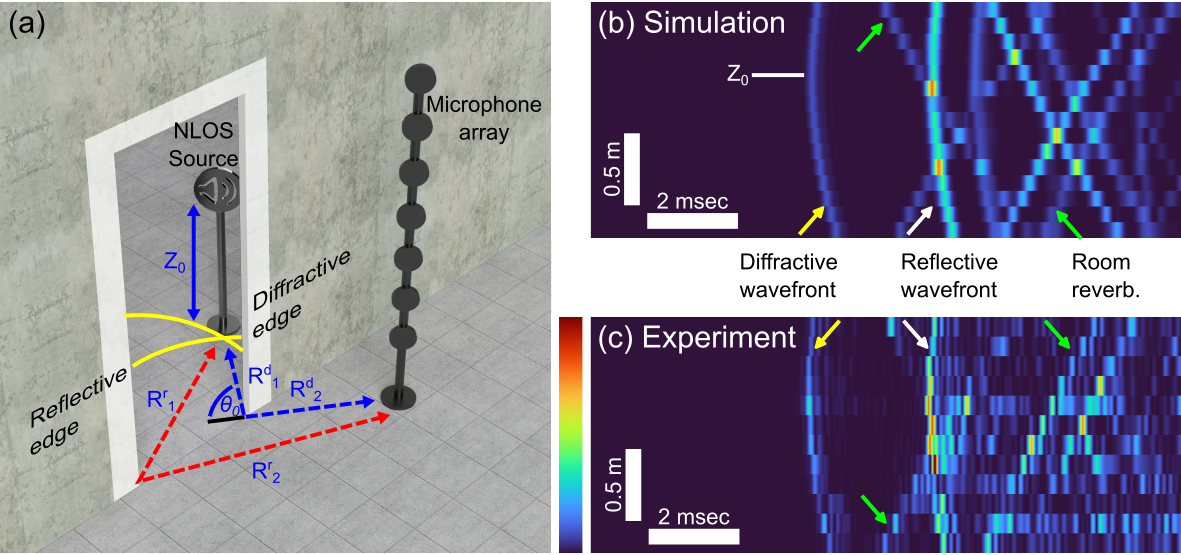}
    % \includesvg{Figures/door_scenario_fig.svg}
    \caption{
    \label{fig:door_scene} (a) Conceptual illustration of NLOS localization in a doorway scenario. (b-c) The amplitude envelope of the signal detected by the microphone array in this scenario, $A(z,t)$.
    The detected signals from the hidden acoustic point source are the result of two phenomenae (depicted by arrows):
    The spherical wavefront that interacts with the door edge that is closer to the microphone array undergoes diffraction, which redirects the wave toward the microphone array. This diffracted wavefront (yellow arrows) is the first to be detected by the array.
    The spherical wavefront that interacts with the farther edge of the door is reflected towards the microphone array. This reflected wavefront (white arrows) is the second to be detected by the array.
    Additionally, later arriving reverberations echo throughout the room (green arrows).
    Colorbar range in (b-c) is [0-1].
    }
\end{figure*}

\begin{figure}[!ht]
    \centering
    \includegraphics{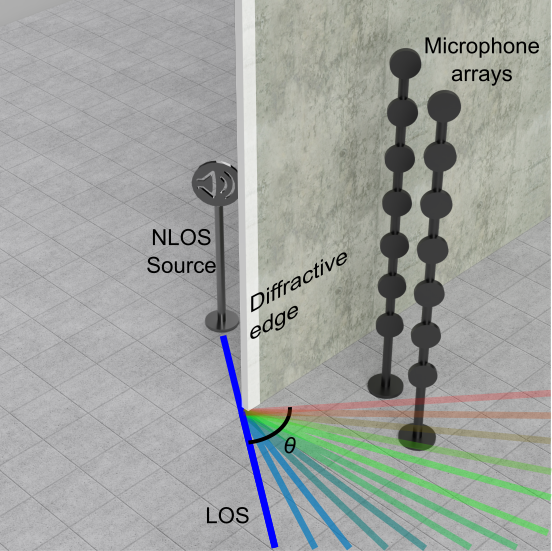}
    % \includesvg{Figures/KED_scenario_fig.svg}
    \caption{
    \label{fig:ked_scene} Conceptual illustration of NLOS localization in a knife-edge scenario. The spherical wavefront propagates from the hidden point source and interacts with the edge, resulting in diffraction toward the microphone array.
    The diffraction loss introduced by this interaction is dependent on both the angle $\theta$ between the microphone array and the line-of-sight (LOS), and on the signal frequency (depicted by different colors). Lower-frequencies diffract more efficiently at high azimuths.
    }
\end{figure}

Here, we consider two common scenarios of passive acoustic NLOS localization without a relay surface: a doorway, and a single edge. In the doorway scenario (Fig.~\ref{fig:door_scene}a), the two edges of a doorway act effectively as two virtual line-arrays, where the signal from one edge is relayed through diffraction and from the other edge through both diffraction and reflection. These two virtual arrays can be utilized as the conventional 2D relay surface used in previous works. In contrast, as we discuss below, in the case of a single edge that acts as a single virtual line-array (Fig.~\ref{fig:ked_scene}), additional information is required to assess the azimuth of the source.

In this work, we address passive acoustic NLOS source localization in these two scenarios. Using an array of microphones, we measure the acoustic field on the detection side that originates from a hidden point source emission.
In the case of a doorway (Fig.~\ref{fig:door_scene}a), we calculate the TOA of the measured signals at two visible edges of the door. The TOA from the two line-arrays is then used as in a traditional NLOS TOA reconstruction\cite{wetzstein2020NLOS, velten2012NLOS3d} to localize the source in 3D.
In a single-edge scenario, as is encountered in a single barrier edge (Fig.~\ref{fig:ked_scene}) or a convex corner, we similarly utilize the calculated TOAs at the single edge, which allow us to localize the range and height of the source. 
The azimuthal angle, $\theta$, is determined from the relative spectral signature of the knife-edge diffraction\cite{hum2007ked_theory} measured by two microphone arrays, exploiting the fact that lower frequencies diffract more efficiently at large diffraction angles. 
The use of a spectral signature to determine location is similar to the human perception of 3D acoustic location via the head-related transfer function (HRTF)\cite{wenzel1993HRTF, cheng2001intro_HRTF, bruschi2024review_HRTF}.

%##### DOOR METHOD FIGURE #####

% ======================
%##### SCENARIO I - 2 edges, a door #####  
% Wavefront types detected in a door scenario
We first consider the doorway scenario, as illustrated in Fig.~\ref{fig:door_scene}a. In this scenario a hidden source emits a short acoustic pulse. Part of the acoustic waves from the source are diffracted in a knife-edge diffraction from the closer edge of the door to a microphone array (dashed blue curve in Fig.~\ref{fig:door_scene}a), and another part is reflected from the farther edge of the door to the microphone array (dashed red curve). The challenge is to utilize these two received waves to localize the hidden source in 3D.

\begin{figure*}[!ht]
    \centering
    \includegraphics[scale=1.1]{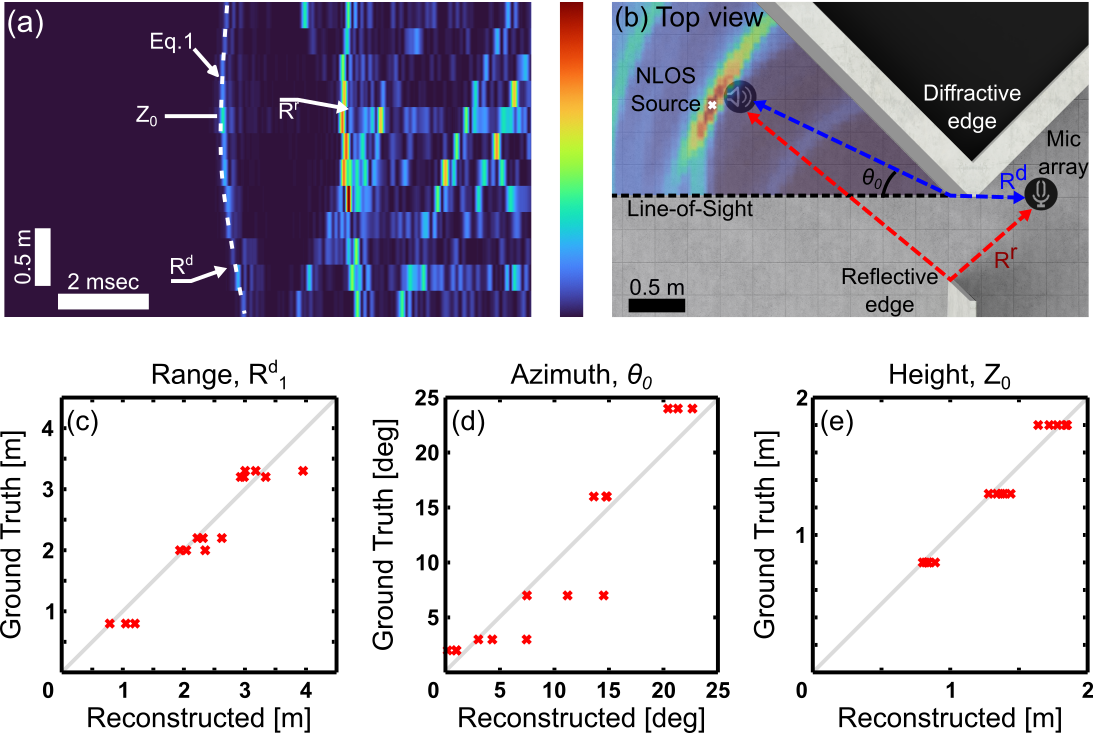}
    % \includesvg{Figures/door_recon_fig.svg}
    \caption{
    \label{fig:door_recon_fig}
    (a) The amplitude envelope of the signal detected by the microphone array in a doorway scenario, $A(z,t)$. The spherical wavefront attributes of Eq.~\ref{eq:spherical_wf} are depicted on the measured signal.
    (b) A top view of the scene and the localization heatmap reconstructed from the data in (a).
    (c) The reconstructed range between the source and the microphone array shows an RMSE of approximately $18\%$.
    (d) The reconstructed azimuth angle between the LOS and the microphone array shows an RMSE of approximately $3^{\circ}$.
    (e) The reconstructed height of the source shows a mean prediction error of approximately $6\%$.
    Colorbar range in (a-b) is [0-1].
    }
\end{figure*}

We begin our study with a numerical analysis of this scenario using an FDTD simulator (k-Wave \cite{treeby2010kwave1, treeby2018kwave2}) to simulate the propagation of acoustic fields in the scene, and the resulting signals expected to be measured by the microphone array. The simulated 3D scene geometry is depicted in Fig.~\ref{fig:door_scene}a (and in a top view in Fig.~\ref{fig:door_recon_fig}b). It consists of a vertical array of 15 microphones placed on a rod at a distance of $R_2^d=0.8m$ from the diffractive edge of a doorway. Walls, ceiling, and floor were simulated as high speed of sound and density compared to air (
%Different positions of the hidden source are simulated on the other side of the door.
see Supplementary Information for full simulation parameters).

One example result of the simulated waves at the microphone array for an object located at $R_1^{d}=3.2m, \theta_0=25^\circ, z_0=1.5m$ is presented in Fig.~\ref{fig:door_scene}b. The plotted values are the envelope amplitude after bandpass filtering at 500Hz-9KHz and amplitude demodulation, at each microphone height, $z$, over time, $t$: $A(z,t)$. 
As expected, the received signals are comprised of two types of waves and reverberations. The three types waves are depicted by arrows in Fig.~\ref{fig:door_scene}b,c: 
the diffracted wave from the closer edge of the doorway (yellow arrow), 
the reflected wave from the farther edge of the doorway (white arrow), 
and reverberating waves following each of the above (green arrows) that originate from multiple reflections from other surfaces or objects in the scene.

Considering a point source, we expect the detected wavefront to be spherical, where the radius of curvature is defined by the propagation distance from the source, i.e. $R^{d}_{1}+R^{d}_{2}$, and $R^{r}_{1}+R^{r}_{2}$. The diffracted wavefront has a radius of curvature that is shorter than the reflected wavefront radius of curvature (See Fig.~\ref{fig:door_recon_fig}a,b). 
Explicitly, $R^{d}_{1}+R^{d}_{2}\equiv R^{d} ~<~ R^{r}\equiv R^{r}_{1}+R^{r}_{2}$, where $R^{d}_{2}$ and $R^{r}_{2}$ are known scene-constants (propagation distances from the edges to the detection array, Fig.~\ref{fig:door_scene}a), and $R^{d}_{1}$ and $R^{r}_{1}$ are the unknown parameters of the hidden source location.

% The detected spherical properties of a wavefront
% Reverberations can be observed as mirrored point sources located outside of the room, beyond its walls, floor and ceiling. Consequently, they reach the microphone array at a relatively large angle, therefore are detected with a smaller effective aperture. As these wavefronts propagation distance is relatively greater than the effective aperture, they seem like plane-waves at detection.
% On the contrary, a point source located inside the room is detected with a larger effective aperture, therefore allowing the capture of the spherical properties of the detected wavefront.
% add a figure in Supp. for this explanation? Explain by addressing the effective-apperture / range ?

Ignoring the reverberations background, in each microphone, each of the two wavefronts (diffracted of reflected) is manifested as a single time of arrival (TOA). For a hidden point source located at a height $z_0$ above the floor and at a propagation distance ($R^r$ or $R^d$) from the microphone array, emitting a short pulse at an unknown time $t_0$, the wavefront time of arrival, $TOA(z)$, is given by:

\begin{equation}
    \label{eq:spherical_wf}
    TOA(z) = t_0 + \sqrt{(z-z_0)^2 + R_0^2}
\end{equation}

where:
(I) $R_0$ will be equal to either $R^{d}$ or $R^{r}$ for either the diffracted or reflected wave, and will be manifested in the curvature of the detected wavefront.
(II) The height of the source above the ground, $z_0$, which will be also the height where the detected signal will first arrive (see $z_0$ mark in Fig.~\ref{fig:door_scene}b).
(III) The unknown emission time, $t_0$, which %In this manuscript, we observe the case where the transmission time is unknown. $t_0$, 
causes a temporal shift of the detected waves.
Note that $R_0$ also causes a temporal shift of the detected waves, but unlike $t_0$ will also change the detected wavefront curvature.%, while a change in $t_0$ will yield only a shift along the time-axis.
%In other words, a late TOA can be either because the source is far (a large $R_0$), or because it fired late (a large $t_0$), but the two are resolvable. % NOT SURE HOW NEEDED IS THIS LAST SENTENCE...

%##### DOOR RESULTS FIGURE #####

% fitting data to spherical waves and reconstructing the scene
We have used Eq.~\ref{eq:spherical_wf} to localize the source in 3D in three steps: First, we found the TOA of the diffracted wavefront in each microphone. This was done by finding the first peak of the signal that was significantly stronger than the mean background amplitude.

Secondly, we fitted the found TOA of the diffracted wavefront to Eq.~\ref{eq:spherical_wf} (Fig.~\ref{fig:door_recon_fig}a), yielding $R^d$, $t_0$, and $z_0$. 

%Following this step, the scene is reconstructed using the two edges as a 2D rely surface, as described below.
% To detect the TOA of the first wavefront in each of the detection channels, traditional methods of thresholding can be used.
% The TOA of all of the channels form a wavefront. The found wavefront can be fitted to a spherical one to assess the three attributes above. These parameters allow us to infer the height of the source ($z_0$) and the transmission time ($t_0$) with an accuracy of approximately $6\%$ and the sampling rate, respectively. Nevertheless, even after restricting the observed height to a specific plane, this method provides the range relative to the diffractive edge ($R_{d}$) alone, which is insufficient for localization.

% using the ranking method to find the most probable location
Knowing $z_0$ and $t_0$ from the initial fit, we can now use a modified beamforming algorithm using the two edges as a 2D rely surface, to reconstruct the initial amplitude of the source at each point in the hidden scene, as we explain below. An example of the beamforming reconstructed amplitude metric from one experimental measurement is presented as a heat-map in Fig.~\ref{fig:door_recon_fig}b. The location of the source is found from the maximum amplitude metric location.
%This is done under the assumption that the geometry of the visible side is measurable.
% The needed distances are: $R^{d}_1$ and $R^{r}_1$ that are the distances between the microphone array and the obstacle's edges at $z=0$ (floor-level), the aperture's width, and the location of each microphone in the array.

% Each location on the hidden side of the scene is fully-defined by its distance to the diffractive and to the reflective edge of the obstacle. Therefore, each location $\bm{r}$ can be described as $\bm{r}=r\left(R^{d}, R^{r}\right)$.
% This allows linking each point in the hidden scene to two curves in the spatio-temporal display of the detected signal's envelope, $E(z,t)$, as can be observed for the detected signal from the point-source in Fig.~\ref{fig:door_recon_fig}a.

The reconstruction heatmap is generated according to the following: 

The metric, $M$, for each point is calculated using the summation of the absolute-value of the signal's envelope under the mask's restrictions:

(I) For each spatial position in the reconstructed grid $(x,y,z)$, the horizontal propagation distances to the microphone array $R^{d,r} = \sqrt{(x-x_{d,r})^2+(y-y_{d,r})^2}+R_2^{d,r}$ are calculated. 

(II) For each of the two distances $R^d,R^r$, Eq.~\ref{eq:spherical_wf} is used to provide two TOAs for all microphones: $TOA_r(z)$, $TOA_d(z)$.

(III) For each of the microphones the amplitude in the trace $A(z,t)$ at a temporal gate of length $0.2ms$ around each of the two TOAs is averaged, providing two scalar numbers that provide an estimate for the detected amplitudes of the reflected and diffracted waves: $A_r$ and $A_d$, respectively. 

%corresponding to the propagation distance from the two edges: $E(z,t(z;R^{d}))$ and $E(z,t(z;R^{r}))$. 

(IV) A metric, $M$, for the current point is calculated by:

\begin{equation}
    \label{eq:rank_scene}
    M\left(R^{d},R^{r}\right)= 
    \frac{A_d}{(R^d)^2} \frac{A_r}{(R^r)^2}
\end{equation}

% results of the reconstruction
where the multiplicative weighting by $\left(1/R^dR^r\right)^2$ was chosen to prioritize closer locations, and to reduce the effect of strong late-arriving reverberations.
Fig.~\ref{fig:door_recon_fig}b displays the reconstructed metric from one experimental measurement performed as explained below.

The above reconstruction is in fact a delay-and-sum beamforming of the detected amplitudes, ignoring the detected phases, i.e. equivalent of using the two edges of the door as a the relay surface. We have empirically found that it gave better results to conventional delay-and-sum on the measured datasets. %  metric is similar to propagating the detected signal to each of the edges and then generating a beamformed image from each edge. 
%The scene rank will be, similar to incoherent-compounding imaging, the product of the absolute-value of the two beamformed images.

Following the numerical study, we performed a set of experiments in a large concrete-walled basement with the geometry given in Fig.~\ref{fig:door_recon_fig}b. As the hidden source, we used hand claps of a test subject at various positions hidden outside the line of sight behind the doorway (colored area in Fig.~\ref{fig:door_recon_fig}b). 
The spectrum of the hand-claps was significant above the background noise level mostly in the frequency range of $500Hz-9kHz$ (Supplementary Fig.~S1). 
The signals were recorded by a vertical array of 15 microphones with $13cm$ spacing between neighboring microphones, starting at a height of $46cm$ above the floor, up to $228cm$ above the floor. The distance of the array from the diffractive edge was $0.8m$, and the width of the doorway was $0.9m$.
15 different source locations were tested. For each test, a single hand clap was measured, and the position of the source was reconstructed as explained above. The results for the reconstructed positions vs. the real source positions are presented in Fig.~\ref{fig:door_recon_fig}c-e. Each of the panels presents one coordinate of the source position: its distance from the diffractive edge ($R^d_1$), its azimuth relative to the LOS ($\theta_0$), and its height ($z_0$). 

%Ea source was positioned at a variety of ranges ($R^{d,r}$) beyond the LOS, 
%The reconstruction method was applied to the measured signals for a localization of the point-source in each scene. 

Note that the reconstructed location is sensitive to errors in the estimation/measurement of the geometry of the visible part of the scene. In particular, the measured distance between the array and the edges ($R^{d}_2$ and $R^{r}_2$) causes an additive constant bias to the azimuth. The presented results in Fig.\ref{fig:door_recon_fig}b are obtained after a small change of $2cm-8cm$ in the measured scene parameters, which is within our experimental measurement error. %Therefore, assuming the error should be non-biased, we've modified the geometric measurements up to a reasonable tolerance to minimize the mean error.

Inspecting the results, for our experimental parameters, the reconstructed range root-mean-squared-error (RMSE) is approximately $18\%$ of the true range.
%This is likely due to the fact that when sampling with a limited aperture, spherical wavefronts of different curvatures are harder to differentiate as the radius of curvature gets larger. 
The reconstructed azimuth RMSE is approximately $3^{\circ}$, and the reconstructed height RMSE is approximately $6\%$, which is at worst less than the $13cm$ spacing between the microphones.

% ====================================
% ====================================
%##### KED SCENE FIGURE #####

%#####  SCENARIO 2 - 1 edge, KED #####
Following the completion of the empirical study of the doorway scenario, we turned to consider the scenario of NLOS localization in a semi-infinite aperture when only a single edge is visible, as illustrated in Fig.~\ref{fig:ked_scene}, as is the case when the source is hidden beyond e.g. a corner of a building. In this case, the signal arriving at the microphone array will only include the diffracted wavefront (and the latter reverberations). Thus, while the diffracted wavefront measured in a single vertical array can provide the range and height of the source, additional information is required to estimate its azimuth $\theta_0$.
In our results below, we estimate the azimuth from the spectral ratio of the diffracted waves as measured by two vertical arrays (Fig.~\ref{fig:ked_scene}). 
The two arrays allow us to exploit the fact that lower frequencies have stronger diffraction amplitude at large angles, resulting in an azimuth-dependent spectrum (depicted as colored lines in Fig.~\ref{fig:ked_scene}).

Theoretically, the diffracted amplitude from a semi-infinite aperture, also known as knife-edge diffraction, is dependent on the hidden source azimuth, $\theta$, the frequency, $f$, and the geometry of the scene \cite{hum2007ked_theory}. These parameters are encompassed in a single parameter, $\nu$, termed the Fresnel-Kirchoff parameter \cite{hum2007ked_theory}: $ \nu \propto \theta \sqrt{f}$.

The diffraction amplitude is commonly characterized by the ratio between the detected diffracted NLOS signal ($A_\textit{knife-edge}$) and the signal that would have been detected at the same position in a free-space configuration ($A_{free-space}$). This ratio is termed the 'diffraction loss' (or gain), $\mathscr{L}$ \cite{hum2007ked_theory}.
To provide an intuitive insight into the expected result, we review the common simple case of knife-edge diffraction in 2D under the Fresnel approximation (See Supplementary Material Section B). For this simplistic case, $\mathscr{L}$ can be numerically calculated using the complex Fresnel integral\cite{hum2007ked_theory}:

\begin{equation}
\label{eq:ked_loss_vs_nu}
 \mathscr{L}(\nu) \equiv   \frac{A_\textit{knife-edge}}{A_\textit{free-space}}  = \frac{1+j}{2} \int_\nu ^\infty {\exp{\left(-j\frac{\pi}{2}t^2\right)}dt}
\end{equation}

%Where $\nu$ is called the Fresnel-Kirchoff parameter. This parameter depends on some of the scene's geometries, and is proporional to the source's azimuth times the square-root of frequency:

%\begin{equation}
%\label{eq:nu_def}
   % \nu \propto \theta \sqrt{f}
%\end{equation}

\begin{figure*}[!ht]
    \centering
    \includegraphics{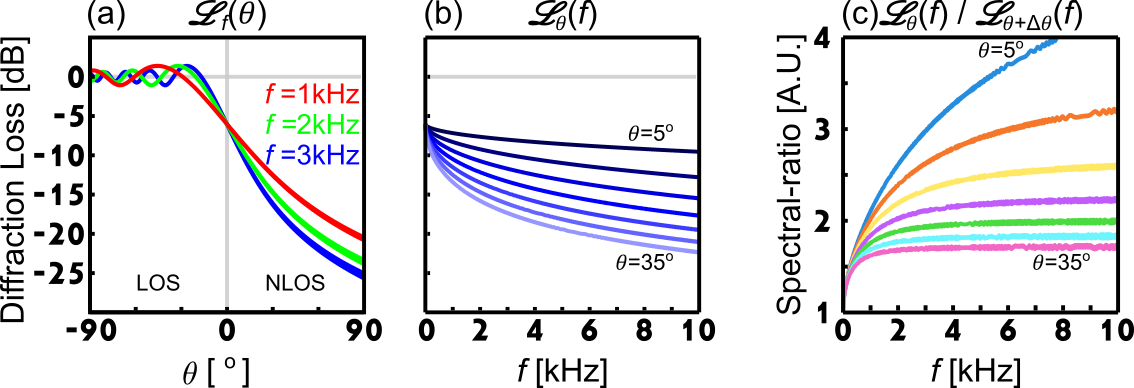}
    \caption{
    \label{fig:ked_loss_fig}
    Theoretical knife-edge diffraction loss for a simplistic 2D scenario (See Supplementary Material Section B).
    (a) Knife-edge diffraction introduces a loss that depends on the diffraction angle ($\theta$, as defined in Fig.\ref{fig:ked_scene}) and the frequency. This panel shows numerical calculations of the induced loss in a specific scenario.
    (b) In the frequency domain, the knife-edge loss exhibits a unique spectral signature corresponding to the source’s azimuth $\theta$. This panel displays numerical calculations of the spectral signature for a hidden source located at azimuths ranging from $5^{\circ}$ to $35^{\circ}$.
    (c) Direct measurement of the loss spectrum without including the source’s spectrum is not possible. To isolate the loss effect, two microphone arrays positioned at an angular distance of $\Delta \theta=25^\circ$ between them are used to measure the diffracted signal. The spectral ratio between the two arrays reveals the impact of the diffraction-loss on the spectrum, independently. This panel shows numerically calculated spectral ratios for a hidden source at azimuths ranging from $5^{\circ}$ to $35^{\circ}$.
    }
\end{figure*}

Fig.~\ref{fig:ked_loss_fig}a,b presents the theoretical diffraction loss according to Eq.~\ref{eq:ked_loss_vs_nu} of the simple 2D case, for different frequencies and azimuths. 
As expected, the diffraction loss is larger at larger angles (Fig.~\ref{fig:ked_loss_fig}a) and at higher frequencies (Fig.~\ref{fig:ked_loss_fig}b).
% angle ($\theta$) is larger, o (depicted by line transparency in Fig.~\ref{fig:ked_scene}). Also, for larger frequencies (depicted by blue-hued lines in Fig.~\ref{fig:ked_scene}), the induced loss will be larger, causing them to diminish first as the diffraction angle gets larger.

Fig.~\ref{fig:ked_loss_fig}b shows the dependence of the measured NLOS spectrum on the hidden source azimuth, $\theta$. In the case that the emission spectrum at the source position is known, a single measurement of the diffracted spectrum can yield the source azimuth using Fig.~\ref{fig:ked_loss_fig}b.
%the measured spectral  , one can argue that sources of different azimuths will undergo a different signature of spectral-loss. 
Alas, in many realistic scenarios, the source emission spectrum is not known. Thus, the spectral loss is unattainable from a single measurement as the measured spectrum is the product of the unknown emitted spectrum and the diffraction loss. 
To overcome this limitation, we measure the diffracted NLOS signal with \textit{two} microphone arrays positioned at different azimuths from the diffractive edge (Fig.~\ref{fig:ked_scene}). We found that, in both theory and experiments, the spectral ratio between the two arrays allows the retrieval of the diffraction-loss while cancelling any dependency on the source spectral signature (Fig.~\ref{fig:ked_loss_fig}c). Intuitively, the ratio of the spectral amplitudes measured in two microphone arrays at different azimuths gives, for each frequency, $f$, the ratio between the diffraction losses at these two azimuths: $\mathscr{L}_\theta(f) / \mathscr{L}_{\theta+\Delta \theta}(f)$.

% Analogous to HRTF
This method is analogous to the human perception of an acoustic source localization via spectral signatures using the head-related transfer function (HRTF) \cite{wenzel1993HRTF, cheng2001intro_HRTF, bruschi2024review_HRTF}. 
While determining if a source is located to the left or the right of the listener is rather straightforward by measuring the relative TOAs (and/or amplitudes) of the signals received in the left vs. the right ear, determining the height of a source or if it is located in front or at the back of the listener cannot be determined from TOAs alone. However, the spectral transfer function that arises from the propagation of the signal from a source location to the ear is dependent of the source height due to diffraction from the earlobes, head, and shoulders. This set of head related transfer functions (HRTFs) affects the spectral signature of each source and allows acoustic localization in 3D \cite{wenzel1993HRTF, cheng2001intro_HRTF, bruschi2024review_HRTF}.

% experiments
To validate our approach, we performed a set of experiments in a large concrete-walled basement with a convex corner. As the hidden source, we used hand claps of a test subject at various positions hidden outside the line of sight. 
The spectrum of the hand claps was significant above the background noise level, mostly in the frequency range of $500Hz-9kHz$ (Supplementary Fig.~\ref{fig:app_spectrums}). 
The signals were recorded by two vertical arrays of 8 microphones each (Fig.~\ref{fig:ked_scene}), with a spacing of $26cm$  between neighboring microphones, starting at a height of $46cm$ above the floor, up to $228cm$ above the floor. The distance of both of the arrays from the diffractive edge was $0.8m$, and their azimuths were different by $25^\circ$.
Three source azimuths were tested, all at the same range and height ($R_1^d=3.1m, z_0=1.3m$). The tested azimuth relative to the LOS, $\theta_0$, were $5^\circ, 10^\circ, 15^\circ$. 
For each position, 10 hand claps were measured.
The rather small azimuthal angles were chosen as the spectral ratio metric is more noise-sensitive and tolerance-sensitive for larger azimuths (above $~20^\circ$,  Fig.~\ref{fig:ked_loss_fig}c).

To optimally filter out background noises and room reverberations, we used a short temporal window on the received signals around the first arriving wavefront to calculate the spectral ratio of the first-arriving wavefront alone. We found that the spectral-ratio is sensitive to the exact choice of window size. In our experiments the window size was fine-tuned until optimal results were produced, at a window-size of $0.7usec$.

\begin{figure}[!ht]
    \centering
    \includegraphics{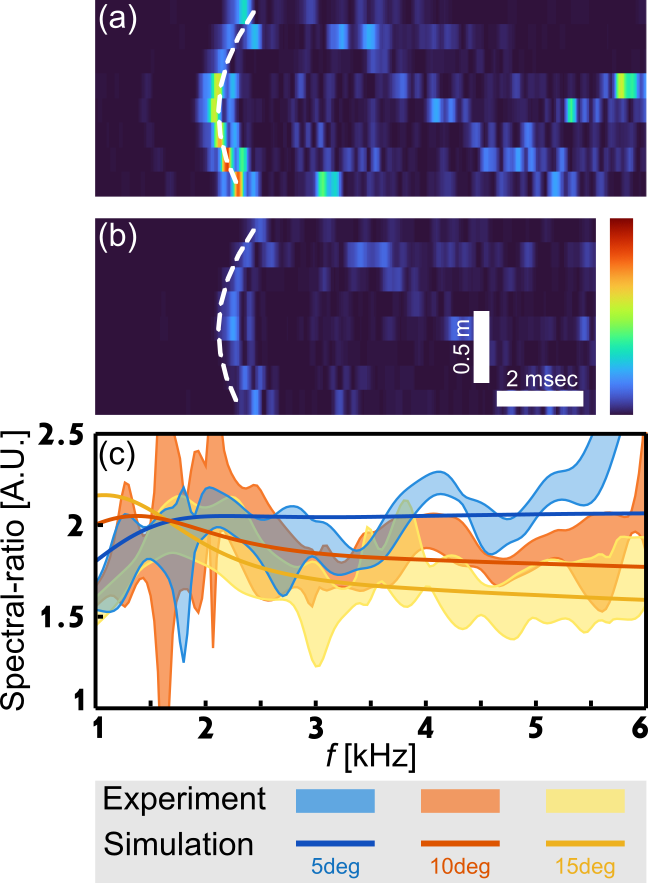}
    \caption{
    \label{fig:ked_results}
    (a,b) Experimental amplitude envelope of the signals recorded by the two microphone arrays in a knife-edge scenario experiment (Fig.~\ref{fig:ked_scene}). Dashed lines indicate the detected first arriving wavefronts, whose arrival times are used for retrieving the height and distance of the source. The spectral ratio between the first arriving wavefronts in the two arrays is used to determine the source azimuth (c).
    (c) Spectral ratios of the first arriving wavefronts between the two microphone arrays. Color areas display the experimental  one standard-deviation across several measurements. Solid lines: numerically simulated spectral ratio obtained from an FDTD simulation of acoustic wave propagation in the scene.
    Colorbar range in (a-b) is [0-1].
    }
\end{figure}

Fig.~\ref{fig:ked_results}a,b presents an example of the amplitude envelope of a single measured signal at the two arrays, $A_{1,2}(z,t)$. Dashed white curves are the fits for the first arriving wavefronts from the measured TOA. The height and distance of the source were determined from the two curves.
The azimuth information was assessed from the spectral ratio, $\mathscr{L}_\theta(f) / \mathscr{L}_{\theta+\Delta \theta}(f)$, calculated directly by the ratio between the power spectra of the signals in the temporal window around the first arriving wavefront. For each measurement, the power spectrum in each pole was calculated from the average of the power spectra of the three central microphones whose height is closest to the source height.

% After a temporal filtering of the first-arriving wavefront, the spectral-ratio was calculated for each pair of similar-height microphones by dividing the spectrum of the low-azimuth microphone by the higher-azimuth microphone. 

% explaining the area plots
Fig.~\ref{fig:ked_results}c displays the experimental spectral ratio for each of the three tested azimuths as colored areas. Each colored area represents the range of one standard-deviation of the measured spectral ratios in the experimental measurements of one azimuth. For each azimuth, the spectral-ratio was calculated for all valid measurements, where we have disregarded 
measurements of hand claps that resulted in a relatively low intensity, an undetectable wavefront, or a wavefront fit whose RMSE is larger than 0.1msec.

Fig.~\ref{fig:ked_results}b compares the experimentally measured spectral ratios (colored areas) to numerically simulated spectral ratios (solid-line plots), obtained by FDTD simulation of acoustic waves propagation in the scene, using k-wave \cite{treeby2010kwave1, treeby2018kwave2} (See Supplementary Material Section A).
As expected, at high frequencies, where the spectral ratio is expected to be more significantly different between different azimuth (Fig.~\ref{fig:ked_loss_fig}c), the experimental spectral ratios are indeed close to the numerically simulated ones (Fig.~\ref{fig:ked_results}c, solid lines), while displaying also additional spectral modulations, which we attribute to interferences in the scenes, such as those originating from reflections from the body of the hand-clapping human subject.

%##### KED RESULTS FIGURE #####

%##### DISCUSSION \ CONCLUSION #####
In conclusion, in this work we proposed an approach for NLOS localization without a relay surface utilizing edge diffraction for passive acoustic NLOS localization.
The new capability allows to address scenarios where traditional methods, which rely on reflections from visible relay surfaces, cannot address, as we demonstrate in the single-edged convex corner scenario.

Our numerical results provide a foundation for the proof-of-principle experiments. However, advancements in several areas are required to improve the robustness and applicability of these diffraction-based techniques. In particular, while we have put forward a framework for the analysis of the spectral-ratio based information in the single-edged convex corner scenario (Fig.~\ref{fig:ked_results}c), a more advanced signal processing algorithmic approach, e.g. utilizing neural networks, is likely to improve performance. 
Our experiments highlighted the sensitivity of the reconstructed locations to accuracy of measuring the visible scene geometry, and the sensitivity to the processed temporal window size. 
Future work on more advanced algorithms that would reduce these sensitivities is important for deployment in uncontrolled environments.

Important next steps include exploring more complex, dynamic, and noisy environments. %Real-world scenarios often involve significant reverberations from multiple surfaces and objects in the scene, which were observed in both simulations and experiments alongside the primary diffracted and reflected waves.
Additionally, utilizing non-impulsive emissions, and emmissions from distributed sources would be interesting extensions to explore.

Overall, our work demonstrates the potential of acoustic diffraction as a valuable source of information for passive NLOS localization, opening interesting avenues for sensing in complex environments. 
%Building upon these foundations in future work can address current limitations and unlock the full potential of this technique.

% ---------------------------------------
 
\begin{acknowledgments}
This project was supported by the H2020 European Research Council (101002406).
\end{acknowledgments}

\section*{Author Contributions}

\textbf{Tal Sommer:}
Conceptualization (equal);
Data Curation (lead);
Formal Analysis (lead);
Methodology (lead);
Writing/Original Draft Preparation (lead);
Writing/Review \& Editing (equal).
\textbf{Ori Katz:}
Conceptualization (equal);
Supervision (lead);
Funding Acquisition (lead);
Writing/Review \& Editing (equal).

\section*{Data availability}
The data that support the findings of this study are available from the corresponding author upon reasonable request.

\section*{Supplementary Material}
See supplementary material for the simulation parameters and a comparison of a hand-clap with the experimental background noise.

\bibliography{PANL}

\end{document}

\title{Passive acoustic non-line-of-sight localization without a relay surface}
\author{Tal I. Sommer}
%\email[]{tal.sommer@mail.huji.ac.il}
\affiliation{Department of Applied Physics, Hebrew University of Jerusalem, Jerusalem 9190401, Israel}
\affiliation{Alexander Grass Center for Bioengineering, Hebrew University of Jerusalem, Jerusalem 9190401, Israel}

\author{Ori Katz}
%\email[Author to whom correspondence should be addressed: ]{orik@mail.huji.ac.il}
\affiliation{Department of Applied Physics, Hebrew University of Jerusalem, Jerusalem 9190401, Israel}

\date{\today}
\maketitle

%%%%%%%%%%%%%%%%%%%%%%%%%%%%%%%%%%%%%%%%%%%%%%%%%%%%%%%%%%%%%%%%%%%%%%%%%%%%%%%%%%%%%%%%%%%%%%%%%%%%%%%%%%%%%%%%%%%%%%%
% SUPPLEMENTARY 1
\clearpage{}
%\begin{multicol}{1}

\subsection{\label{app:SIM_PARAMS}K-wave Simulation Parameters}

All simulations (detected signal in Fig.~1b and detected signal used for spectral-ratios calculations in Fig.~5c in the main text) were performed in MATLAB 2025a, using k-Wave, an open source MATLAB toolbox designed for the time-domain simulation of propagating acoustic waves \cite{treeby2010kwave1, treeby2018kwave2}.
K-wave toolbox is propagating an initial given pressure field through heterogeneous media by iteratively solving the continuity equation (conservation of mass) and Euler’s equation (conservation of momentum), assuming a linear adiabatic equation of state.

Scene geometry in the simulations was designed according to the experimental scenes, as described in the main text. Walls, floor, and ceiling were all simulated as highly reflective: $c_{air}=343m/s, ~\rho_{air}=1.29kg/m^3, ~c_{reflectors}=3c_{air},  ~\rho_{reflectors}=10\rho_{air}$. 3 times the velocity and 10 times the density, relative to air, was enough to simulate high reflectance with no numerical artifacts.
Air pores were inserted in the obstacle wall to make sure as much of the signal is reflected.
The grid was of cubed $16mm$ voxels, with an extent of $(D_x, D_y, D_z) = (600,600,200)$ pixels.
Time steps were defined to be of 1usec.

To simulate an impulsive source, similar to a hand clap, the sound source emission was defined as a $40\%$ cycle of 5kHz.

In the doorway scenario (Fig.~1b), the microphone array was simulated as a linear array of 15 microphones, equally spaced between $46cm$ height from the floor level and $228cm$ (pitch $13cm$). The array's distance from the diffractive-edge was $0.8m$, and the door's width was $0.9m$.
In the knife-edge scenario (Fig.~5c), the signal was measured at two microphone arrays beyond the LOS (Fig.~3), positioned at a relative angle of $\Delta \theta =25^\circ$ between them. The microphone arrays were simulated as two linear arrays, each of 8 microphones equally spaced between $46cm$ height from the floor level and $228cm$ (pitch $26cm$). The array's distance from the diffractive-edge was $0.8m$.

There was no addition of noise to the simulation.

The frequency response of the microphone array was defined as a gaussian filter centered around 5kHz with a bandwidth of 200\%.
Signal was demodulated with a rectangular bandpass filter between $500Hz-9kHz$.

%%%%%%%%%%%%%%%%%%%%%%%%%%%%%%%%%%%%%%%%%%%%%%%%%%%%%%%%%%%%%%%%%%%%%%%%%%%%%%%%%%%%%%%%%%%%%%%%%%%%%%%%%%%%%%%%%%%%%%%
% SUPPLEMENTARY 2
\subsection{\label{app:2D_SIM_PARAMS}Fresnel Integral Simulation Parameters}

We review the common simple case of knife-edge diffraction in 2D under the Fresnel approximation.
A simulation is used to numerically asses the diffraction-loss (Eq.~3, and plots in Fig.~4 in the main text). Calculations were performed in MATLAB 2025a.

Scene geometry in the simulation was designed according to the experimental scenes, simplified to a 2D case, as described in the main text.
The detector was placed at a distance of $0.8m$ from the diffractive-edge, and the source was placed at a distance of $3.2m$, on the other side of the occluder.
Speed-of-sound was assumed to be $343m/s$, as in air.

Spectral-ratio in Fig.~4c was obtained by dividing the diffraction-loss of two azimuths, spaced $25^\circ$ and at the same distance from the edge.

%%%%%%%%%%%%%%%%%%%%%%%%%%%%%%%%%%%%%%%%%%%%%%%%%%%%%%%%%%%%%%%%%%%%%%%%%%%%%%%%%%%%%%%%%%%%%%%%%%%%%%%%%%%%%%%%%%%%%%%
% SUPPLEMENTARY 3
\renewcommand{\thefigure}{S\arabic{figure}}

\subsection{\label{app:spectrum}Source and Noise Spectrum}
\setcounter{figure}{0}

\begin{figure}[!ht]
    \centering
    \includegraphics{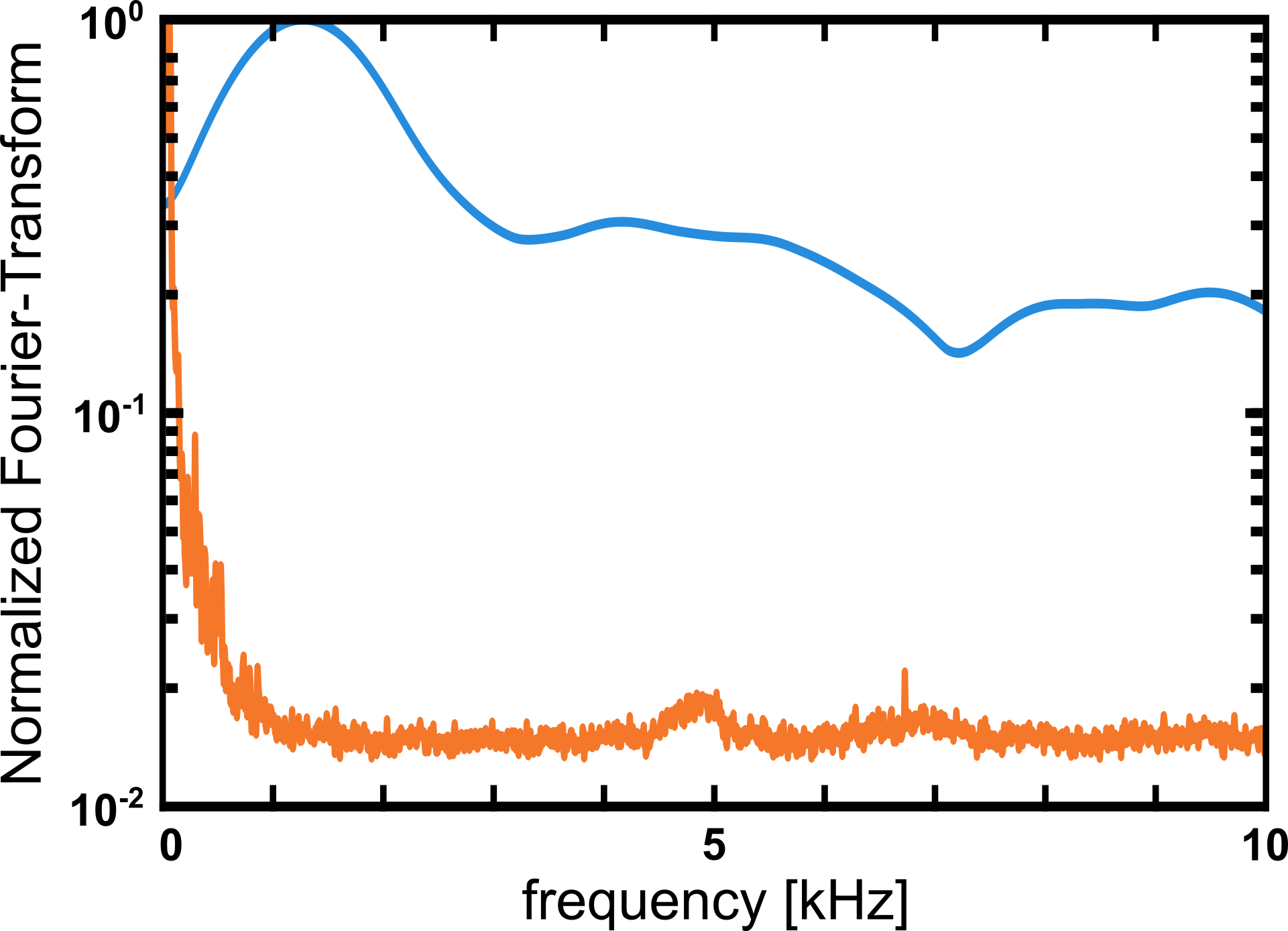}
    \caption{
    \label{fig:app_spectrums}
    The average spectrum of a hand-clap (blue), compared to the average spectrum of background noise (orange), both averaged over 160 measurements. One can see that the spectrum of the hand-claps was significant above the background noise level, mostly in the frequencies above $300Hz$, and peaked at the frequencies below $3kHz$.
    }
\end{figure}

%\end{multicol}{1}
\bibliography{PANL}